\documentclass[fleqn,10pt]{wlscirep}
\title{Control of automated guided vehicles without collision by quantum annealer and digital devices}

\author[1,2*]{Masayuki Ohzeki}
\author[3]{Akira Miki}
\author[1]{Masamichi J. Miyama}
\author[3]{Masayoshi Terabe}
\affil[1]{Graduate School of Information Sciences, Tohoku University, Sendai 
980-8579, Japan}
\affil[2]{Institute of Innovative Research, Tokyo Institute of Technology, Kanagawa, 226-8503, Japan}
\affil[3]{Electronics R \& I Division, DENSO CORPORATION, Tokyo 103-6015, Japan}

\affil[*]{mohzeki@tohoku.ac.jp}

\begin{abstract}
We formulate an optimization problem to control a large number of automated guided vehicles in a plant without collision.
The formulation consists of binary variables. A quadratic cost function over these variables enables us to utilize certain solvers on digital computers and recently developed purpose-specific hardware such as D-Wave 2000Q and the Fujitsu digital annealer.
In the present study, we consider an actual plant in Japan, in which vehicles run, and assess efficiency of our formulation for optimizing the vehicles via several solvers.
We confirm that our formulation can be a powerful approach for performing smooth control while avoiding collisions between vehicles, as compared to a conventional method.
In addition, comparative experiments performed using several solvers reveal that D-Wave 2000Q can be useful as a rapid solver for generating a plan for controlling the vehicles in a short time although it deals only with a small number of vehicles, while a digital computer can rapidly solve the corresponding optimization problem even with a large number of binary variables.
\end{abstract}
\begin{document}

\flushbottom
\maketitle
%
%
\thispagestyle{empty}

\section*{Introduction}
An automated guided vehicle (AGV) is a portable robot for moving materials in manufacturing facilities and warehouses \cite{Ullrich2014,Fazlollahtabar2015,Fazlollahtabar2016}.
It moves along markers or wires on floors or uses vision, magnets, or lasers for navigation in a few cases. 
Currently, in most of plants, transportations of materials relies on AGVs and their smooth control.
However, in limited-size plants, AGVs are frequently involved in traffic congestion around intersections because a large number of AGVs cross them simultaneously.
AGVs are controlled by following a predetermined rule to prevent collisions between AGVs.
The simplest but most important rule prohibits AGVs from crossing intersections simultaneously.
In other words, an intersection is not allowed to be occupied by multiple AGVs.
One approach for designing a plant that is suitable for controlling AGVs might be to divide an area into several zones and prohibit simultaneous occupation by multiple AGVs in an event period \cite{Li2011,Li2015,Li2016}.
In addition, the scheduling problem for controlling the AGVs is also considered \cite{Fazlollahtabar2015}.

In the present study, we consider a method for controlling AGVs without changing the layout of a plant.
The method is based on the formulation of the control problem of AGVs as an optimization problem.
We determine the routes of AGVs without collision by solving the optimization problem.
In the optimization problem, we utilize binary variables, which represent the selection of routes from numerous candidates in a discretized map. We employ a quadratic cost function with the binary variables to formulate the optimization problem for simplicity.
It nevertheless requires considerable time to solve the optimization problem with binary variables in most cases.
We implement specialized solvers to provide meaningful solutions for controlling AGVs.
One of the solvers is D-Wave 2000Q, which solves the unconstrained binary quadratic programming problem (recently also termed as the quadratic unconstrained binary optimization (QUBO) problem) using quantum annealing (QA) \cite{Kadowaki1998}.

Quantum annealing utilizes the quantum effect for solving the QUBO problem \cite{Lucas2014}.
The quantum effect introduces a driving force in searching for the minimum of a cost function.
The mechanism is purely quantum, but it can be mapped onto a stochastic dynamical system in which binary variables fluctuate.
The fluctuation is used to search for the ground state, which corresponds to the minimum of the cost function.
Quantum annealing can find the ground state at the end of its protocol by gradually decreasing the strength of the fluctuation of binary variables.
Theoretical assurance is established by following the quantum adiabatic theorem \cite{Suzuki2005,Morita2008,Ohzeki2011c}.
Surprisingly, QA is realized in an actual quantum device using present-day technology \cite{Dwave2010a,Dwave2010b,Dwave2010c,Dwave2014}. 
Since then, QA has been developed rapidly and has attracted considerable attention. It has been tested for numerous applications such as portfolio optimization \cite{Rosenberg2016}, protein folding \cite{Perdomo2012}, the molecular similarity problem \cite{Hernandez2017}, computational biology \cite{Richard2018}, job-shop scheduling \cite{Venturelli2015}, traffic optimization \cite{Neukart2017}, election forecasting \cite{Henderson2018}, and machine learning \cite{Crawford2016,Arai2018nn,Takahashi2018,Ohzeki2018NOLTA,Neukart2018,Khoshaman2018}.
In addition, its potential might be boosted by the nontrivial quantum fluctuation, referred to as the nonstoquastic Hamiltonian, for which efficient classical simulation is intractable \cite{Seki2012,Seki2015,Ohzeki2017,Arai2018dy}.
However, the solution is not always optimal owing to the limitations of devices and environmental effects \cite{Amin2015}.
Theoretical assurance is given under the assumption of the nonexistence of an environment.
This is not a realistic situation in performing QA in quantum devices such as D-Wave 2000Q.
Therefore, several protocols based on QA do not follow adiabatic quantum computation or maintain a system in the ground state to reach the optimal solution in the final stage of QA; rather, they employ a nonadiabatic counterpart \cite{Ohzeki2010a,Ohzeki2011,Ohzeki2011proc,Somma2012}.
In addition, even though the outputs from D-Wave 2000Q are not optimal solutions, we may use these as approximate solutions.
The approximate solutions can be at a satisfactory level depending on the purpose of the solution of the QUBO problem.

In the present study, we formulate the QUBO problem for controlling AGVs and solve it via D-Wave 2000Q.
In the control of AGVs, rapid response is necessary for dealing with instantaneous changes in a system.
Thus, it is expected that D-Wave 2000Q can provide a method for establishing the future infrastructure for controlling AGVs because it can output approximate solutions in a few tens of microseconds.
We also utilize other solvers for attaining the approximate solutions of our QUBO problem
and compare their performance with D-Wave 2000Q. In addition, we discuss the potential applications of D-Wave 2000Q and its future development.

The remaining part of the paper is organized as follows:
In the next section, we formulate the control of AGVs as the QUBO problem, which can be solved using D-Wave 2000Q.
The solution does not always satisfy certain constraints for controlling AGVs, and output solutions must be postprocessed.
We explain how to attain reasonable solutions via the postprocessing.
In the third section, we solve the QUBO problem via D-Wave 2000Q and its original linear programming problem via the Gurobi Optimizer \cite{gurobi2018}.
In the following section, we compare the results obtained by D-Wave 2000Q and other solvers.
In the last section, we summarize our study and discuss the direction of future work on the optimization problem in terms of the QUBO problem and its solvers.

\section*{Methods}
We formulate the optimization problem for controlling AGVs in this section.
The formulation is generic and not specific to individual situations.
We do not optimize the entire plan to control all AGVs simultaneously.
We consider iterative optimization to provide an adequate route for each AGV during time period $T$.
At time $t_0$, we gather information on the location, $x_i$, and the task, $s_i$, distributed to each AGV.
We solve our optimization problem and employ its solution to control the AGVs during time period $T$.
After moving the AGVs at $t_0 + T$, we again gather information on the current situation and iterate the above procedure.

Let us formulate the optimization problem to be solved for a plan in time period $T$.
We define the binary variable for each AGV as $q_{\mu, i} = 0,1$, where $\mu$ is the index for a route and $i$ is that for an AGV.
The index of the route is selected from a set of routes, $M(x_i,s_i)$, where $s_i$ is the given task for the $i$-th AGV.
The index of $i$ runs from $1$ to $N$, which is the number of AGVs.
The set of routes is constructed a priori following the tasks and the structure of the plant in which the AGVs run.
Then, we define the cost function for controlling the AGVs as
\begin{equation}
f_0({\bf q}) = - \sum_{i=1}^N \sum_{\mu \in M(x_i,s_i)} d_{\mu} q_{\mu,i},
\end{equation}
and the following constraints are satisfied: 
\begin{equation}
\sum_{\mu \in M(x_i,s_i)} q_{\mu,i} = 1 \quad {\forall} i
\end{equation}
and
\begin{equation}
\sum_{i=1}^N \sum_{\mu \in M(x_i,s_i)} F_{\mu,t,e} q_{\mu,i} = 1 \quad {\forall} t,~{\forall}e.
\end{equation}
where $t$ ranges from $t=1$ to $t=T$ and $e$ denotes an edge in the routes for $F_{\mu,t,e} \neq 0$ in the plant.
The cost function is optimized to maximize the length of the routes of the AGVs.
The first constraint is to enforce each AGV to select a single route.
The second constraint is to avoid collision.
This constraint is created by defining a binary quantity for characterizing the $\mu$-th route as $F_{\mu,t,e}$ with $0$ and $1$.
For each route, $F_{\mu,t,e}=1$ on the edge occupied by the selected route, $\mu$, at time $t$.
On the contrary, $F_{\mu,t,e}=0$ on the edge unoccupied by the selected route, $\mu$, at time $t$.
In a previous study \cite{Neukart2017}, a similar formulation was proposed for the traffic optimization of moving taxis.
However, the study did not consider the time dependence of $F_{\mu,t,e}$.
In the present study, the speed of the AGVs is almost constant.
The AGVs can move as expected and can be predicted precisely.

The above optimization problem is a linear programming problem, and thus, it can be straightforwardly solved using a conventional method.
However, by employing the penalty method, we now formulate the linear programming problem as
\begin{equation}
f({\bf q}) = - \sum_{i=1}^N \sum_{\mu \in M(x_i,s_i)} d_{\mu} q_{\mu,i} + \lambda_1 \sum_{i=1}^N \left( \sum_{\mu \in M(x_i,s_i)} q_{\mu,i} -1 \right)^2 + \lambda_2 \sum_{e \in E}\sum_{t=1}^T \left( \sum_{i=1}^N \sum_{\mu \in M(x_i,s_i)} F_{\mu,t,e} q_{\mu,i} - 1\right)^2,\label{QUBO}
\end{equation}
where $E$ denotes all edges of the network along which the AGVs move in the plant, and $\lambda_1$ and $\lambda_2$ are predetermined coefficients.
The second and third terms represent the constraints of the original linear programming problem.
Thus, we do not allow any solution with a nonzero value in the these terms.
To attain solutions that satisfy these constraints, we must set relatively large values of coefficients $\lambda_1$ and $\lambda_2$ based on the penalty method.
We here make a comment on the relationship of our problem with the previous study for reducing the traffic flow in the literature \cite{Neukart2017}.
We expand the third term in Eq. \label{QUBO} and then obtain $\lambda_2 \sum_{e \in E}\sum_{t=1}^T \left( \sum_{i=1}^N \sum_{\mu \in M(x_i,s_i)} F_{\mu,t,e} q_{\mu,i}\right)^2 - 2\lambda_2 \sum_{i=1}^N \sum_{\mu \in M(x_i,s_i)} d_{\mu} q_{\mu,i} + {\rm Const.}$, because $\sum_{e \in E}\sum_{t=1}^T F_{\mu,t,e} = d_{\mu}$.
When $\lambda_2=1$, the first term vanishes with the resultant linear term and then the cost function (\ref{QUBO}) coincides with that in the previous study \cite{Neukart2017}.
In this sense, our problem is a generalization of the reduction problem of the traffic flow.
The cost function can be reduced into the following quadratic form with a  matrix $Q$ as
\begin{equation}
f({\bf q}) = {\bf q}^{\rm T} Q{\bf q} + {\rm Const.}.
\end{equation}
where ${\bf q}$ is the vector consisting of binary variables $q_{\mu, i}$.
Thus, the above optimization problem is transformed into the unconstrained binary quadratic programming or QUBO problem.

In order to shorten the time of the whole procedure to control the AGVs, we prepare a database that stores the set of routes when we create the QUBO during the time period $T$. 
In advance, we generate the shortest paths from an origin to a destination for each task.
We divide the shortest paths into sets of several vertices at the longest $vT$, where $v$ is the maximum speed of the AGVs, and store them. 
When we build the QUBO matrix, we only elucidate a vertex set included in a part of the shortest paths beginning at $x_i$ up to the reachable position at the end of period $T$.
For instance, let us consider the case that the part of the shortest path for the $i$-th AGV at $x_i=1$ consists of the node set $\{1,2,3\}$.
The shortest path is determined by the initial point, $1$, and the destination for accomplishing the task, $s_i$. This results in $\{1,2,3\}$.
Then, we prepare the route sets as $\{1\}$, $\{1,2\}$, and $\{1,2,3\}$, which indicate "stop," "$1$ step ahead," and "$2$ steps ahead," respectively.
This approach is similar to a classical method for collision avoidance between AGVs in that it introduces delays and deviations in the shortest paths for each vehicle \cite{Dowsland1994}.

Next, we describe how to solve the QUBO problem efficiently.
We use D-Wave 2000Q to solve the QUBO problem.
A binary variable is regarded as a spin through an interpretation as $2q_{\mu, i}-1 = \sigma^z_{\mu, i}$.
In QA, the spin represents quantum-mechanical operators known as the $z$-component Pauli matrices as $\sigma_{\mu,i} \to \hat{\sigma}_{\mu, i}$, where $\hat{\dot}$ denotes the quantum-mechanical operators that form the quantum system.
In QA, the spin represents quantum-mechanical operators known as the $z$-component Pauli matrices as $\sigma_{\mu,i} \to \hat{\sigma}_{\mu, i}$, where $\hat{\dot}$ denotes the quantum-mechanical operators that form the quantum system.
D-Wave 2000Q exploits quantum fluctuation, which is frequently termed as the transverse field, to attain the solution of the QUBO problem through the quantum dynamics governed by the following Hamiltonian \cite{Kadowaki1998}:
\begin{equation}
\hat{H}(s) = A(s) f(\hat{{\boldsymbol \sigma}}) + B(s) \sum_{(\mu,i)} \hat{\sigma}_{\mu,i}^x
\end{equation}
where $\hat{\sigma}_{\mu,i}^x$ is the $x$-component of the Pauli matrices to generate the superposition of the up ($\sigma^z_{\mu,i}=1$) and down ($\sigma^z_{\mu,i}=1$) spins.
Starting from the trivial initial state with full superposition at $s=0$, quantum dynamics leads to a nontrivial ground state, which corresponds to the minimum of the cost function at $s=1$, where $A(s)$ and $B(s)$ are time-dependent parameters defined as $A(s=0)=B(s=1)=0$ and $A(s=1)=B(s=0)=1$.
D-Wave 2000Q performs QA on its superconducting qubits.
D-Wave 2000Q does not always output the optimal solution at the end of its protocol because it is not the device for performing ideal QA\cite{Amin2015}.
The ideal QA is assumed to perform in an isolated quantum system.
In fact, D-Wave 2000Q typically outputs low-energy states, namely, approximate solutions with relatively low-valued cost functions.
In addition, annealing time is extremely short owing to its limitation caused by the coherence time of superconducting qubits.
Therefore, we can attain numerous outputs from D-Wave 2000Q for the same QUBO problem in a short time.
This is a weak but outstanding feature of the D-Wave 2000Q.
The feature is weak because solutions do not satisfy the constraints as outputs are not always optimal.
In other words, in a few cases, nonzero values remain in the second and third terms of Eq. \ref{QUBO} for each output obtained from D-Wave 2000Q.
However, the solutions employed to control the AGVs must satisfy all constraints.
We utilize the remarkable advantage of D-Wave 2000Q, which enables us to attain a large number of outputs as the candidates of good solutions and not necessarily optimal solutions.
We may filter out the solutions that do not satisfy the constraints from the outputs of D-Wave 2000Q.
As a result, we obtain a reasonable solution without collisions and the double selection of routes.
 
We may utilize other solvers for finding the minimum of the QUBO problem.
In the present study, we test the Fujitsu digital annealer (DA) to solve the QUBO problem \cite{Tsukamoto2017,Aramon2018}.
Fujitsu Laboratories has recently developed purpose-specific CMOS hardware designed to solve fully-connected QUBO problems (i.e., on complete graphs). This is an advantage over D-Wave 2000Q, in which the chimera graph is employed.
The DA hardware is currently able to solve Ising-type optimization problems with a size of up to $1024$ variables, with $26$ and $16$ bit (fixed) precision for biases and variable couplers, respectively.
In the DA hardware, simulated annealing (SA) \cite{Kirkpatrick1983} with several improvements is employed to solve the QUBO problem.
The improved SA uses a parallel-trial scheme, in which a flip of all variables in parallel is attempted at each step, and seeks higher acceptance probability for updating one-spin flip compared to the ordinary process in SA.
In addition, the DA utilizes an escape mechanism referred to as a dynamic offset to overcome short and narrow barriers.

As described earlier, our QUBO problem is originally the linear programming problem.
In the present study, we test direct manipulation to solve the original linear programing problem using the branch and bound method via the Gurobi Optimizer \cite{gurobi2018}.
In the Gurobi Optimizer, we first solve a relaxed optimization problem by mapping the binary variables to continuous variables.
The solution determines the lower bound of the cost function in the original optimization problem.
Then, we solve the original linear programing problem by the discretization of the solution with continuous variables while reaching the lower bound.
When reaching the lower bound, the tentative solution of the original optimization problem is confirmed to be optimal.
The attained solution is deterministic owing to lack of stochasticity.
This is an essential difference from D-Wave 2000Q and the DA.

Below, we report the comparison among the results obtained by these methods for controlling the AGVs while avoiding collision.
\section*{Results}
\begin{figure}
\begin{center}
\includegraphics[width = 0.5\textwidth]{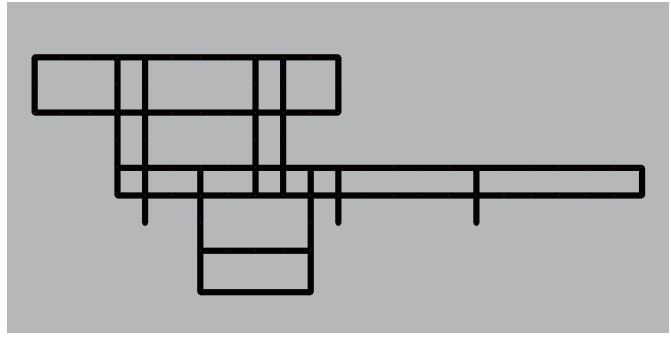}
\caption{Plant used in the present study. In this plant, $10$ AGVs move along a road and complete their tasks.}
\label{map}
\end{center}
\end{figure}

In this section, we report the results attained by iteratively solving the QUBO problem at each time period for controlling the AGVs.
For proving the efficiency of our method, we prepare a simulation environment for an actual plant as shown in Fig. \ref{map}.
The plant utilizes $10$ AGVs for product delivery, and the AGVs move simultaneously along four fixed routes according to predetermined tasks.
The speed of each AGV is $0.5$ m/s.
The distance between nodes is $10$ m.

\begin{figure}
\begin{center}
\begin{minipage}[b]{\linewidth}
\includegraphics[bb = 0 0 691 346, width = 0.7\textwidth]{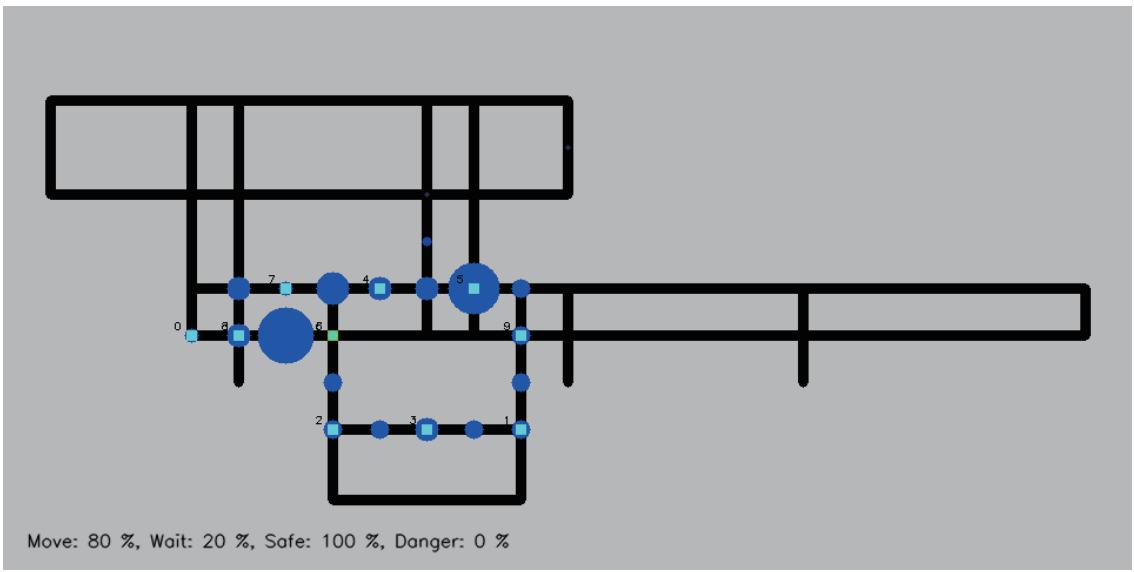}
\end{minipage}
\begin{minipage}[b]{\linewidth}
\vspace{3cm}
\includegraphics[bb = 0 0 691 346, width = 0.7\textwidth]{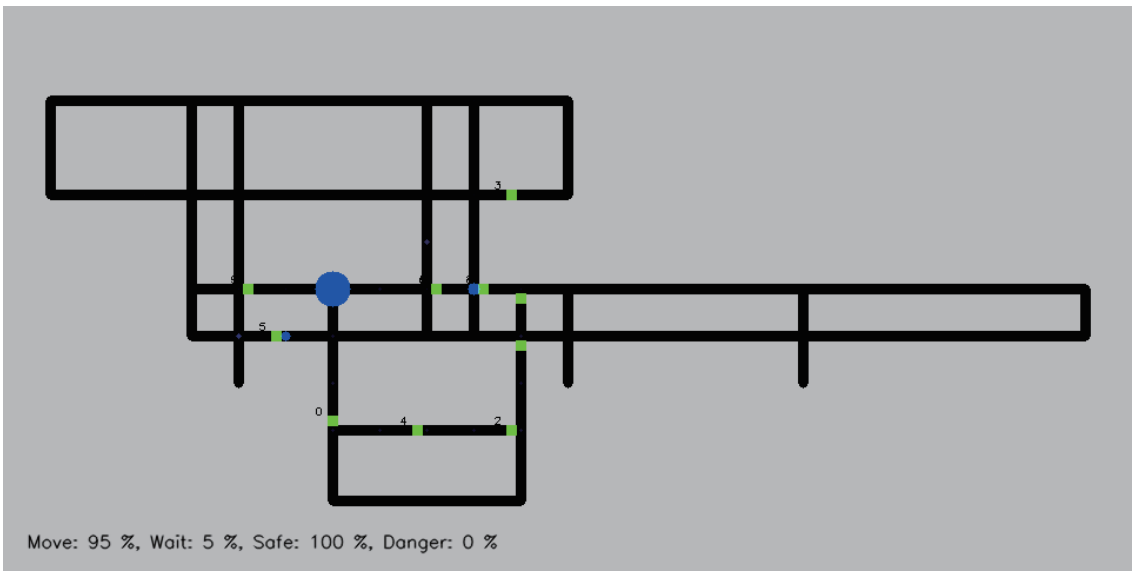}
\end{minipage}
\caption{Comparison among the solvers: (left) Conventional method and (right) D-Wave 2000Q.
The green dots denote the locations of the AGVs at the end.
The blue circles represent the accumulated waiting time for the AGVs.}
\label{map_comp1}
\end{center}
\end{figure}

We simulate the controlled AGV movement using different methods, namely, the conventional method, and the solution of the QUBO problem by D-Wave 2000Q and the DA, and the solution of the original linear programming problem by the Gurobi Optimizer.
The results attained by the conventional method and D-Wave 2000Q are shown in Fig. \ref{map_comp1}.
We simulate the AGVs in the plant for $1000$ seconds and indicate the accumulated waiting time by circles.
The waiting rate is calculated by the ratio of the number of stopping AGVs and the total number of AGVs.
Several circles represent frequent traffic jams of the AGVs.
The size of a circle is proportional to the accumulated waiting time of the AGVs at that point.
The conventional method for controlling the AGVs is a rule-based method at every intersection.

The rule is that when the AGVs require the same intersection route, only one AGV can move in and out at the intersection.
For example, when two AGVs require the same intersection, one AGV waits until the other AGV leaves the intersection. 
The AGVs that move along the circumference of the plant have higher priority for entering an intersection for increasing the working rate.
The time average of the waiting rate converges to $20$ percent. 

We solve the QUBO problem via D-Wave 2000Q at each time period.
The time period is set to be $3$ seconds, namely $T=3$ [s].
We set the parameters as $\lambda_1/(1+\lambda_2) = 1.0$ and $\lambda_2/(1+\lambda_2)=2.0$.
Because D-Wave 2000Q does not deal with large elements of QUBO matrix $Q_{ij}$, the elements of the QUBO matrix is rescaled within the range of the available magnitude.
D-Wave 2000Q solves the QUBO problem $1000$ times for finding reasonable solutions.
We filter the solutions that do not satisfy the constraints and select one of the reasonable solutions for moving the AGVs further.
The AGVs move following the selected solution during the time period of $3$ seconds.
The solution indicates the movement in the next $5$ seconds.
Thus, the movement of the AGVs is updated before they reach the end of the given route.
It can be seen from Fig. \ref{map_comp1} that the number of circles, which represent the accumulated waiting time, is considerably reduced compared to the conventional method.
The time average of the waiting rate converges to $5$ percentages .
The actual movement of the AGVs from the initial condition is shown in the supplemental video files.
Compared to the result of the conventional method, the AGVs move smoothly following the solution to the QUBO problem.
The readers can find the smooth movements of the AGVs in the supplemental movies.

\section*{Other solvers and comparison data}

\begin{figure}
\begin{center}
\begin{minipage}[b]{\linewidth}
\includegraphics[bb = 0 0 691 346, width = 0.7\textwidth]{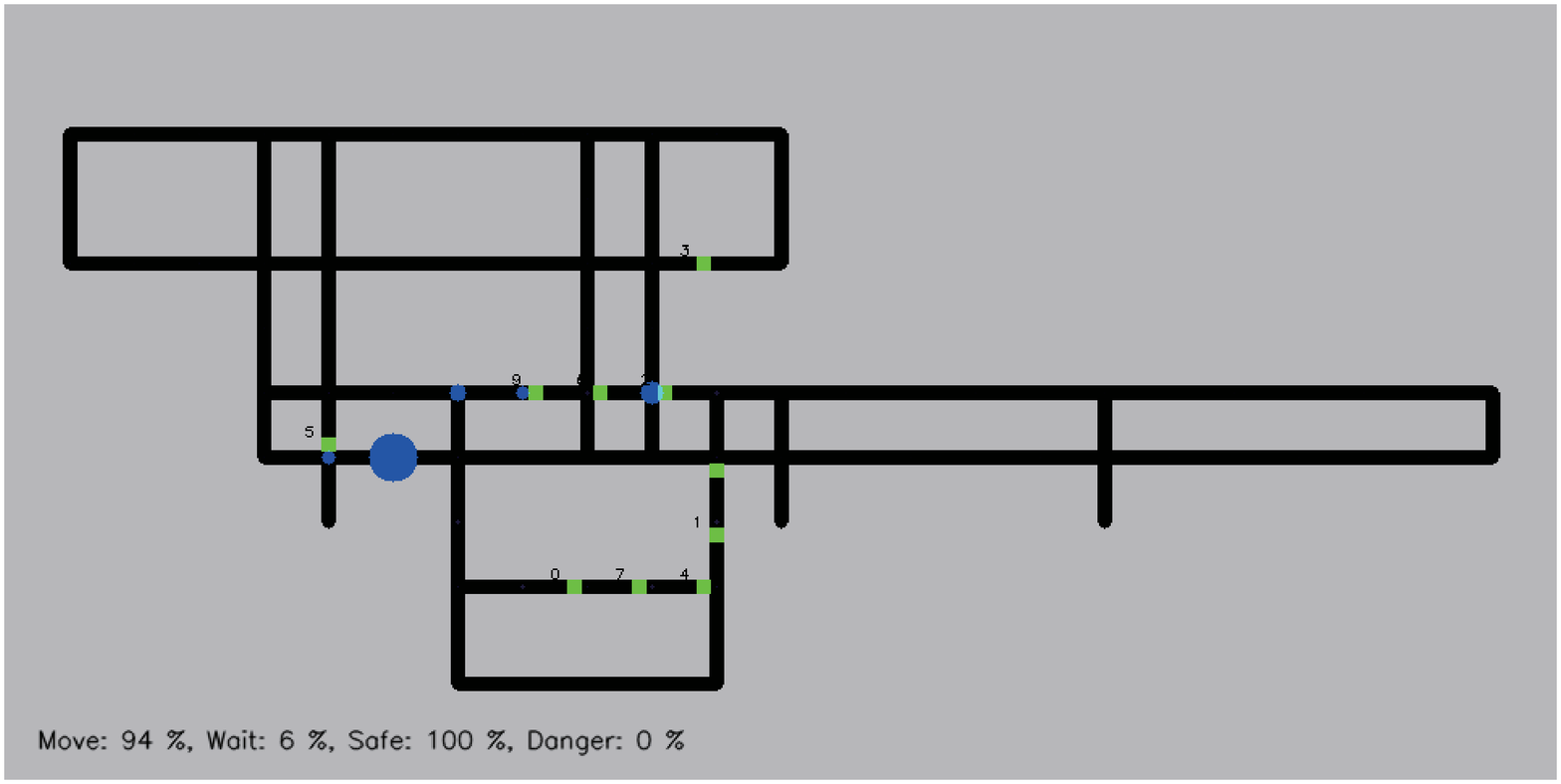}
\end{minipage}
\begin{minipage}[b]{\linewidth}  
\vspace{3cm}
\includegraphics[bb = 0 0 691 346, width = 0.7\textwidth]{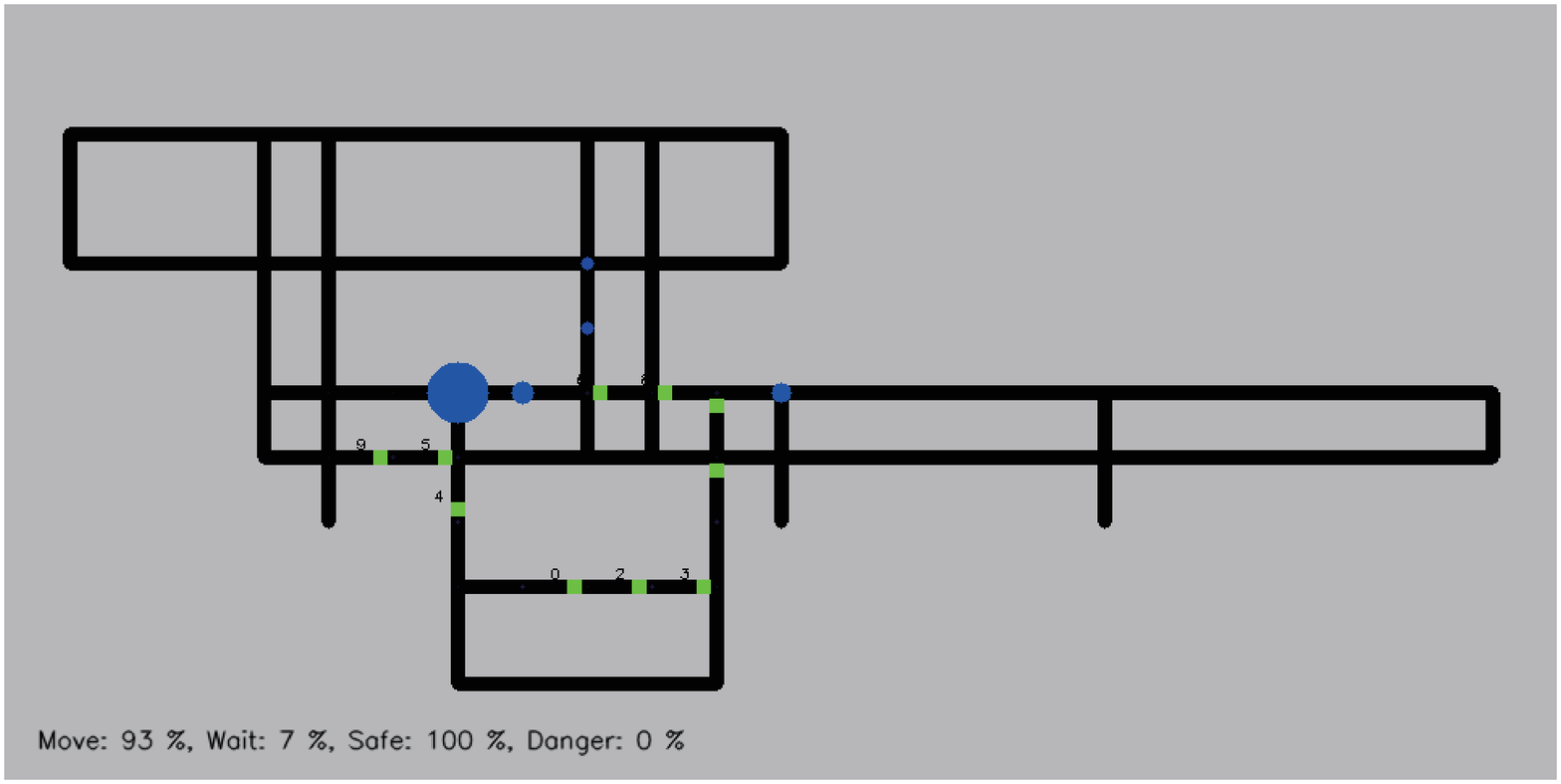}
\end{minipage}
\caption{Comparison among the solvers: (left) Fujitsu digital annealer and (right) Gurobi Optimizer.
The same symbols are used in Fig. \ref{map_comp1}.}
\label{map_comp2}
\end{center}
\end{figure}

It is not necessary to solve the QUBO problem using D-Wave 2000Q; one can utilize other solvers.
One method is the DA, which solves the QUBO problem using an improved version of SA.
Notice that the DA can solve the QUBO problem with a large number of binary variable compared to D-Wave 2000Q.
In the present study, the maximum number of binary variables in the QUBO problem is $60$, which is the product of the number of the AGVs ($10$) and the number of candidates of routes ($6$).
Thus, the number of the binary variables is quite small.
Even though the DA does not exhibit its potential efficiency in this case, we find that the time average of the waiting rate converges around $6$ percent as shown in Fig. \ref{map_comp2}.

In addition, we solve the original linear programming problem through the relaxation of the binary variables to continuous variables by utilizing the branch and bound method via Gurobi Optimizer version 8.01 on a $4$-core Intel i7 4770K processor with $32$ GB RAM.
In this case, we attain the optimal solution of the original linear programming problem in a short time and utilize the optimal solution to control the AGVs.
However, the optimal solution is found through discretization performed using a deterministic rule.
Therefore, the Gurobi Optimizer does not output multiple optimal solutions when the solutions are degenerate.
This deterministic method leads to a biased solution.
The solution attained by the Gurobi Optimizer is the optimal solution for the original linear programming problem. However, it is not always optimal for controlling the AGVs without collisions.
The time average of the waiting rate converges to $7$ percent, which is slightly higher than the results of D-Wave 2000Q and DA.
This is the advantage of the stochasticity of D-Wave 2000Q and the DA, which can search for the ``optimal" solution from a different perspective.
The cost function itself is not necessarily a direct indicator of performance.
Thus the optimal solution for the cost function is not always optimal for the actual performance.
Similar phenomena appear in machine learning.
Generalization performance, which is the measure of potential power in machine learning but not directly related to the cost function to be optimized, can be enhanced via stochastic methods to optimize cost functions.
In particular, QA actually leads to better generalization performance, as shown in literature \cite{Ohzeki2018}.

We repeat the iterative optimization for controlling the AGVs at each time period starting from the same initial condition $10$ times and obtain the average and maximum performance, as shown in Table \ref{working_rates}.
We confirm that D-Wave 2000Q outperforms the other solvers in this problem setting.
The Gurobi Optimizer always leads to the optimal solutions, but the waiting rates are not less than the results obtained by D-Wave 2000Q and DA.
This is because the optimal solutions do not always lead to the best control of the AGVs.
In addition, the biased solution determined by the rule used to find the optimal solution in the Gurobi Optimizer accidentally experiences the worst case scenario for controlling the AGVs.
We can find a better solution in terms of controlling the AGVs by modifying the procedure of finding the optimal solution in the Gurobi Optimizer.
However it is not trivial to find such a modification.
In fact, we add another constraint for the AGVs such that if several AGVs reach the same intersection, the AGV with more following AGVs is preferentially allowed to enter the intersection.
In the formulation of the original linear programming problem, we do not consider the priority with which the AGVs are allowed to enter an intersection in an explicit manner.
We solve the linear programming problem with the additional constraint by employing the Gurobi Optimizer and show its efficiency in Table \ref{working_rates}.
We establish the better cost function and constraints for the Gurobi Optimizer by comparing the solutions obtained by the stochastic solvers and deterministic solver, at least for the present study.
In this sense, the comparison can be beneficial for solving the optimization problem efficiently and formulating a better cost function for accomplishing the actual purpose. 
However, the stochastic search performed by D-Wave 2000Q and the DA is one of the approaches to avoid such intractable fine tuning of the cost function.

\begin{table}[t]
\caption{Working rates of the AGVs obtained by the conventional method, D-Wave 2000Q, Fujitsu digital annealer, Gurobi Optimizer, and modified optimization problem for Gurobi Optimizer.}
\label{working_rates}
\centering
\begin{tabular}{c|ccccc}\toprule[2pt]
 & Conventional & D-Wave 2000Q & Fujitsu digital annealer & Gurobi  & Gurobi +  \\ 
\hline
Average & $80$& $94.2 \pm 1.2$ & $93.4 \pm 1.2$ & $ 93 $& $96$ \\
Max & $80$ & $96$ & $94$ & $93$ & $96$  \\
\hline
\end{tabular}
\end{table}

Below, we discuss the efficiency of the solvers from another point of view, namely the computational time.
We investigate the ``actual" computational time, which is obtained in a standard-user environment, and the quality of the attained solutions against the increase in the number of the AGVs and candidate routes.
We prepare a hundred of different initial locations of the AGVs such as each pair of the AGVs encounter at an intersection and solve the optimization problem.
We report the comparison results in average and variance below.

The D-Wave 2000Q takes $20~\mu$ s, which is predetermined by users, to once solve the optimization problem in the quantum chip with superconducting qubits.
However preprocessing and postprocessing for preparation to solve the optimization problem, the latency of the network when we utilize the D-Wave 2000Q via cloud service, and the queueing time can not be avoided.
Thus the actual computational time takes a little bit longer.
In addition, the D-Wave 2000Q outputs many samples of the solutions once.
We set the number of samples as $1000$ and measure the actual computational time.
We then estimate the actual computational time per output sample as $1.39(33)$ ms for $9$ spins, $1.33(11)$ ms for $21$ spins, $1.51(5)$ ms for $30$ spins, $1.45(12)$ ms for $39$ spins, $1.90(16)$ ms for $51$ spins, and $2.22(22)$ ms for $60$ spins.
These computational times per output sample are only to solve the optimization problems without any assurance of precision of the attained solutions.
The probability for attaining the ground state $P_0$ gradually decreases as the number of spins increases.
In fact, $P_0=1.00$ for $9$ spins, $P_0=0.99(6)$ for $21$ spins, $P_0=0.97(2)$ for $30$ spins, $P_0=0.91(1)$ for $39$ spins, $P_0=0.87(2)$ for $51$ spins and  $P_0=0.74(2)$ for $60$ spins.

The number of spins consists of the multiplication of that of the AGVs and the routes.
The computational time drastically increases for the case of D-Wave 2000Q beyond $60$ spins.
This is due to the limitation of the number of binary variables to be solved simultaneously.
We solve the case with a larger number of binary variables by utilizing qbsolv, which divides the original problem into a number of small problems.
To iteratively use D-Wave 2000Q, we must wait for several seconds owing to the job queue via the cloud service provided by the D-Wave systems Inc. at each iteration to solve the small problems.
The actual computational time per output sample and iteration is $1.80(44)$ ms for $90$ spins, $1.77(59)$ ms for $399$ spins, and $1.37(53)$ ms for $900$ spins.
The iteration numbers become $2 \times 10$ for $90$ spins, $8 \times 10$ for $399$ spins, and $33 \times 10$ for $900$ spins.
The former number in the product is the number of division of the original large problem into small subproblems, and the latter one is that of repetition to solve the optimization problem.  
Thus, the actual computational time can be extremely long.
In addition, the probabilities for attaining the ground state get worse as $P_0=0.27(12)$ for $90$ spins $P_0= 0.03(5)$ for $399$ spins and $P_0=0.001(9)$ for $900$ spins.
This is a weak point to employ the D-Wave 2000Q to solve the optimization problem.
Although it seems that the computational time does not depend on the number of spins, the probability for attaining the ground state gradually decreases as the number of spins increases. 
On the other hand, the Gurobi Optimizer leads to the optimal solutions for each case.
Its computational time to attain the optimal solution depends on the number of spins.
$2.79(6)$ ms for $30$ spins, $3.46(5)$ for $60$ spins $4.25(6)$ for $90$ spins, and $8.70(6)$ ms for $400$ spins.

\begin{figure}
\begin{center}
\includegraphics[width = 0.8\textwidth]{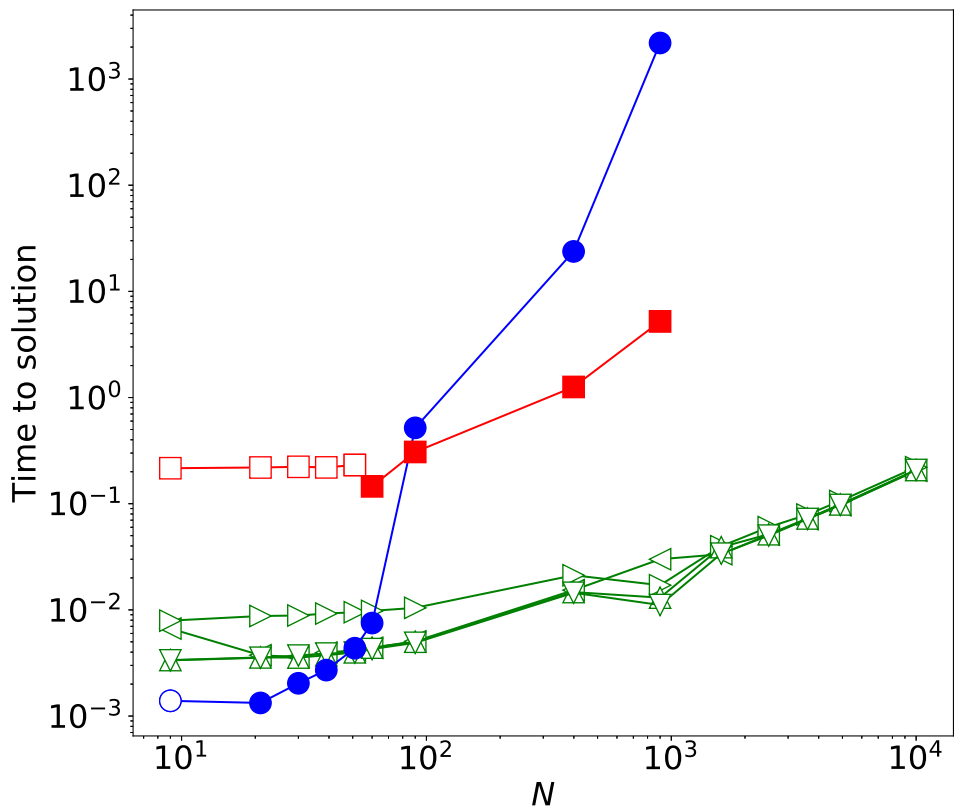}
\end{center}
\caption{Comparison of the time to solution (TTS). 
The filled circles and squares denote the TTS obtained by D-Wave 2000Q and the DA, respectively.
The outlined circles and squares represent the actual computational time (upper bound of the TTS) by the D-Wave 2000Q and the DA.
In addition, we plot the actual computational time by the Gurobi Optimizer by the triangles.
The directions of the triangles distinguish the results by different levels of the ``presolve" option for the Gurobi Optimizer as ``default", ``none", ``conservative", and ``aggressive".}
\label{TTS_fig}
\end{figure}

For the DA, the machine time is set to be enough to solve the optimization problem about $8$ ms.
The actual computational time per output sample takes a little bit longer than the machine time as $0.216(2)$ s for $9$ spins, $0.219(4)$ s for $21$ spins, $0.222(6)$ s for $30$ spins,  $0.220(7)$ s for $39$ spins, and $0.232(9)$ for $51$ spins, $0.240(11)$ for $60$ spins, $0.230(6)$ ms for $90$ spins, $0.336(18)$ s for $399$, and $0.519(32)$ ms for $900$ spins.
Up to $1024$ spins, the current version of the DA can solve once the optimization problem without dividing it into small subproblems.
This is an advantage point of the DA.
In addition, the probability for attaining the ground state $P_0$ is relatively higher compared to that of the D-Wave 2000Q as $P_0 = 1.0$ for $9$, $21$, $30$, $39$ and $51$ spins, $P_0 = 1.000(5)$ fpr $60$ spins, $P_0 = 0.97(3)$ for $90$ spins, $P_0 = 0.71(3)$ for $399$ spins, and $0.37(12)$ for $900$ spins.
Notice that the higher value of the probability for attaining the ground state is obtained by tuning the annealing schedule.
Instead, the actual computational time takes longer.
Then we compute the time to solutions (TTS) defined as
\begin{equation}
{\rm TTS}(p) = t_c \frac{\log(1-p)}{\log(1-P_0)},
\end{equation}
where $t_c$ is the actual computational time per output sample and $p$ is a predetermined precision to attain the ground state.
The time to solution is an indicator of the performance of the solver in the stochastic way.
We show the comparison data of TTS(0.99) of the D-Wave 2000Q and the DA and the actual computational time of the Gurobi Optimizer in Fig. \ref{TTS_fig}.
In the successful cases with $P_0 = 1.0$, we plot the actual computational time instead of the TTS.
The actual computational time per output sample can be upper bound for the TTS.

\section*{Conclusions}
In the setting of the present study, D-Wave 2000Q, the DA, and the Gurobi Optimizer can solve the QUBO problem or the original linear programming problem within the predetermined time period, $T=3$ s.
The number of the binary variables that can be solved within $T=3$ s, which is determined by the product of the numbers of AGVs and routes, are up to $\sim 400$ for D-Wave 2000Q, $\sim 90$ for the DA, and over $10000$ for the Gurobi Optimizer in terms of the TTS and the actual computational time to attain the optimal solution.
Notice that, in order to control the AGVs, it is not necessarily to find the optimal solutions.
The time to solution is just one of the indicators.
As a consequence, for controlling the AGVs via optimization problems, a large  number of degrees of freedom should be dealt with by a digital computer using a modern algorithm such as the branch and bound method.
In contrast, D-Wave 2000Q can be used to obtain a rapid response for controlling AGVs but with small numbers.
In the cases with less than the maximum number of available qubits, D-Wave 2000Q outputs reasonable solutions for controlling AGVs without collision in a short time period.
The Fujitsu DA supports a wide range of applications via the QUBO problem in terms of speed and availability.
In this sense, a digital computer with a modern algorithm and purpose-specific devices, such as D-Wave 2000Q and the DA, are not in conflict.
They each have adequate applications depending on conditions of a problem.
In the present study, we take the case with the AGV speed is relatively slow.
Therefore the time period $T=3$ s is reasonable.

In the present study, we demonstrate the efficiency of the formulation of the optimization problem for controlling the AGVs.
The digital computer with the modern algorithm is the best choice as the solver in terms of the availability and speed in the current situation.
Notice that we employ the actual computational time basically to estimate the performance of the D-Wave 2000Q and the DA, not the machine time.
Therefore, if we can avoid the latency of the communication and queuing time for dealing with the jobs to solve the optimization problem in both of the devices via cloud services, better efficiency can be achieved.
In this sense, the computational time of the D-Wave 2000Q and the DA can be reduced significantly.
For instance, the machine time for optimization problem by the D-Wave 2000Q can be set to be $20 \mu$ s and that of the DA is $8-10$ ms.
The D-Wave 2000Q can be an alternative for solving the optimization problem for controlling the AGVs in real plants.

The time period was set in the present study following the current situation of the real plant, in which several workers walks,
In the cases without any workers, the AGVs can move faster than the setting of the present study.
Then shorter response time for controlling the AGVs is necessary.
The next-generation quantum annealer beyond the D-Wave 2000Q is expected as a candidate for controlling the AGVs in such future plants.
The D-Wave quantum processing units continues to steadily grow in number of qubits.
The precision to find the ground state getting better, the TTS becomes shorter.
Moreover the actual computational time can be also shorter.
In this sense, the shorter response time can be achieved and such future plants can be created by the next-generation quantum annealer, although the current version, the D-Wave 2000, is just a proof of concept and has demonstrated the comparative performance with the modern algorithm on the digital computer.
The present study is the first step toward the efficient control of AGVS in future plants as one of the candidates in the real-world application of the QA.

\bibliography{SR_AGV_ver1}

\section*{Acknowledgements}
The authors would like to thank Shu Tanaka for fruitful discussions, which contributed to the work. 
The present work is financially supported by MEXT KAKENHI Grant No. 15H03699 and 16H04382, and by JST START.

\section*{Author contributions statement}

M.O. conceived and conducted the experiment and analyzed the results.
A. M. created the simulation program for AGVs and implemented it in conjunction with the optimization scheme via D-Wave 2000Q, Fujitsu digital annealer, and the classical solvers.
M. J. M conducted the experiment using D-Wave 2000Q and the Fujitsu digital annealer. M. T. discussed the possibility of the other applications of our method to industry, directed the project in our study, and investigated the possible design of our method.
All authors discussed the details of the results and reviewed the manuscript. 

\section*{Additional information}
Te authors declare no competing interests.

\end{document}